\documentclass[journal]{IEEEtran}

\ifCLASSINFOpdf
\else
\fi

\usepackage{amsmath}
\renewcommand{\wp}{\omega_p}
\renewcommand{\wr}{\omega_r}

\usepackage{widetext}
\usepackage{xcolor}
\usepackage{graphicx}
\usepackage{dcolumn}
\usepackage{bm}
\usepackage{subfigure}
\usepackage{siunitx}
\newcommand{\ket}[1]{\left| #1 \right>} 
\newcommand{\avg}[1]{\langle#1\rangle}

\usepackage[utf8]{inputenc}
\usepackage[T1]{fontenc}
\usepackage{mathptmx}
\usepackage{soul}
\newcommand{\dg}{^\dagger}

\begin{document}
\title{Non-classical Semiconductor Photon Sources Enhancing the Performance of Classical Target Detection Systems}
\author{Haoyu He,
Daniel Giovannini,
Han Liu,
Eric Chen,
Zhizhong Yan,
Amr S.~Helmy,~\IEEEmembership{Senior Member,~IEEE,}
\thanks{The first three authors contributed equally to this work. Haoyu He, Daniel Giovannini,Han Liu, Zhizhong Yan, Eric Chen and Amr Helmy was with the Edward S.~Rogers Department of Electrical and Computer Engineering, University of Toronto, 10 King's College Road, Toronto, Ontario M5S 3G4, Canada}}
\markboth{Journal of Journal of Lightwave Technology, ~Vol.~xx, No.~x, July~2019}%
{Haoyu \MakeLowercase{\textit{et al.}}: Non-classical Semiconductor Photon Sources Enhancing the Performance of Classical Target Detection Systems}
\maketitle
\begin{abstract}
We demonstrate and analyze how deploying non-classical intensity correlations obtained from a monolithic semiconductor quantum photon source can enhance classical target detection systems. This is demonstrated by examining the advantages offered by the utilization of the non-classical correlations in  a correlation based target detection protocol. We experimentally demonstrate that under the same condition, the target contrast obtained from the protocol when non-classical correlations are utilized exhibits an improvement of up to \SI{17.79}{dB} over the best classical intensity correlation-based target detection protocol \cite{lopaeva2013experimental}, under \SI{29.69}{dB} channel loss and excess noise \SI{13.40}{dB} stronger than the probe signal. We also assessed how the strong frequency correlations within the non-classical photon pairs can be used to further enhance this protocol.
\end{abstract}
\IEEEpeerreviewmaketitle
\section{Introduction}
%
%
%
%
\IEEEPARstart{O}{ptical} target detection has been receiving increasing attention owing to many emerging applications in the domains of computing, human/machine interaction, LIDAR, and non-invasive biological imaging, amongst others. Conventionally, the sensitivity of optical target detection could be improved by increasing the source brightness, detector sensitivity or improving the throughput of the optical setup. In addition, it has been shown that such sensitivity could also be significantly boosted through using quantum properties of entangled light  \cite{tan2008quantum,shapiro2009quantum,lloyd2008enhanced}, where one photon serves as a probe and the other as a reference. More recent work proposed that entanglement within non-classical photon pairs could be utilized to enhance the target detection sensitivity even beyond conventional limits encountered in the classical regime \cite{zhang2015entanglement}. However, a significant level of complexity in the instrumentation involved including phase-sensitive joint detection is essential to boost the target detection sensitivity beyond the classical regime limits. As such, formidable challenges lie ahead on the route to harvesting the entanglement advantages because, amongst other issues, it requires sub-wavelength-level stabilization of optical phases between the probe and reference photon. \\

The previous demonstrations that utilize non-classical state of light to enhance target detection systems require table-top, mechanically unstable and poorly scalable setups. For example, previous work has relied entirely on bulk optics and nonlinear crystals, such as BBO and PPLN \cite{zhang2015entanglement,lopaeva2013experimental}, for the generation of the entangled photon pairs that illuminate the target of interest. For practical target detection protocols utilizing non-classical photon pairs, the source and the associated setup, need to offer a form factor which enables both remote operation and quantum state generation in the direct vicinity to the object under illumination.\\

There has been astounding progress in the prowess of non-classical sources in the last decade \cite{kang2016monolithic, kang2015two}. In particular, it has been shown that integrated monolithic semiconductor devices can be used to generate and tailor high-quality quantum states of light in active semiconductor structures, such as AlGaAs devices \cite{valles2013generation, horn2012monolithic}. Such structures can directly produce entangled photon pairs without any additional off-chip interferometry, spectral filtering, compensation, or post-selection and then be coupled effectively into optical fiber or integrated topic target detection systems. The flexibility in waveguide structure design also allows for efficient dispersion control and quantum state engineering.\\

\begin{figure*}
    \includegraphics[width=\textwidth]{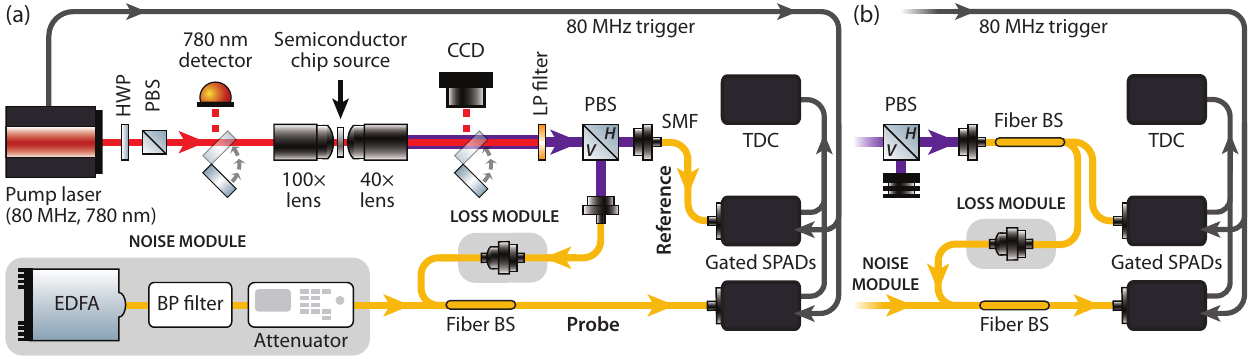}
    \caption{\label{ExpSetup}Schematics of the experimental setup for (a) the nonclassical source enhanced protocol(ICQ), and (b) the classical protocol (ICC). The left-hand side of the ICC setup is omitted for clarity. In ICQ, the H-polarized photon in each pair produced by type-II SPDC is used as a reference beam; the V-polarized photon is used as a probe beam. In ICC, the classically correlated probe and reference beams are obtained by sending the H-polarized photon through a balanced beamsplitter. HWP: half-wave plate; PBS: polarizing beamsplitter; BS: beamsplitter; BRW: Bragg reflection waveguide; LP: long-pass; BP: band-pass; SMF: single-mode fiber; SPAD: single-photon avalanche diode; TDC: time-to-digital converter.} 
\end{figure*}

In this work, we exploit a monolithic quantum light source based on a semiconductor device to enhance the performance of the intensity correlation based target detection protocol, which is otherwise classical. We demonstrate a significant performance enhancement compared to a similar detection system using classical sources. The performance enhancement is also comparable with the previous comparable systems \cite{lopaeva2013experimental} that is based on bulk non-classical light source. In addition, we discussed a possible approach to further enhancing the performance of the intensity correlation based protocols without decreasing the flux of the probe photon, through utilizing the strong frequency correlation within the non-classical photon pairs. 

\section{Intensity correlation target detection with non-classical photon pairs}
\subsection{Target detection with intensity correlation}
The intensity correlation signal could be defined as the covariance between the total photon number operator of the probe mode \(N_p\) and the reference mode \(N_r\) \cite{lopaeva2013experimental}: 
\begin{equation}
S = N_{p}N_{r} - \avg{N_{p}}\avg{N_{r}},
\end{equation}
Note that the average value \(\avg{S}\) is independent of environmental noise and equals zero when the target is absent.
 This property may be useful in practical applications where the noise power is drifting, and a \textit{priori} information about it is hard to obtain\cite{lopaeva2013experimental}. The contrast \(\epsilon\) of the object could be defined as the contrast between \(S_\text{in} \) and \(S_\text{out} \) (the subscript `in' and `out' denote the presence and absence of the object, respectively) and normalized against its standard deviation : 
\begin{equation}
    \epsilon = \frac{\avg{S_\text{in}}-\avg{S_\text{out}}}{\sqrt{\avg{\delta^2 S_\text{in}}+\avg{\delta^2 S_\text{out}}}}\label{SNR_DEF}
\end{equation}

To quantify the sensitivity enhancement brought by the strong intensity correlation of non-classical photon pair sources, we are comparing the intensity correlation target detection protocol with non-classical photon pair sources and that with optimal classical sources, namely, correlated thermal beams \cite{lopaeva2013experimental}. For brevity, these two protocols will be referred to as the ICQ and ICC protocol in the rest of this paper, respectively. However, it should be noted that in practical applications coherent light and intensity detection (will be referred to as the INT protocol) are often used for classical target detection and the ICC protocol may not be optimal. However, the reason why the ICC protocol is considered for comparison is two-fold. First, the ICC protocol has a similar experimental setup and is therefore directly comparable to the ICQ protocol from an implementation point of view. Second, the ICQ protocol we demonstrate here suffers from the large transmission loss of the reference light. This is an experimental imperfection and could be alleviated with a better detector and collection optics. However, since the performance of the ICC protocol is also affected by the transmission loss of the reference photons, comparisons between these two protocols can still reflect the quantum enhancement of the quantum intensity correlation regardless of the experimental imperfection. In the appendix, the ICQ protocol is also compared to the INT protocol. The result shows that the ICQ protocol \textit{cannot} outperform the INT protocol if the INT protocol transmits all the probe light in one single pulse. However, if the INT protocol spreads the probe light in the same number of pulses that have equal average photon number as the ICQ protocol, then the ICQ protocol could possibly outperform the INT protocol, provided that perfect transmission of the reference light can be achieved. 
\subsection{Experimental setup}
The experimental approach for the ICQ protocol relies on the generation of correlated photon pairs from a type-II spontaneous parametric down-conversion (SPDC) process. Our semiconductor quantum light source is capable of generating high-quality entangled states with highly versatile and tunable properties \cite{kang2014two}, including non-degenerate, continuously tunable operation \cite{abolghasem2014widely}. Its operation is based on a dispersion engineered AlGaAs waveguide architecture\cite{horn2012monolithic}.\\

The semiconductor source is pumped using a femtosecond pulsed laser. For each pump pulse, the down-converted photons are always generated with different polarizations in pairs. Therefore the number of vertically polarized SPDC photons and horizontally polarized SPDC photons are always equal. The average number of photon pairs generated by one pump pulse is denoted by \(\mu\), which is typically much less than one. This joint state with correlated photon number in different polarizations could be used to detect the presence or absence of a weakly reflecting object. In the experimental setup for ICQ (shown in Fig.~\ref{ExpSetup}(a)), the signal and idler SPDC photons are deterministically separated via a polarization beamsplitter (PBS) into the reference and probe beams. The reference photon is detected (with total detection efficiency \(\eta_r\), including both optical losses and detection efficiency) immediately and the probe photon is sent toward the reflecting object. The reflectivity of the object is modeled by mixing the probe photons with environment noise upon a fiber beamsplitter with power transmission \(\eta_o\) (\(\eta_o=0\) when the object is absent) \cite{lloyd2008enhanced}. The output of the fiber beamsplitter(the `reflected probe light from the object' and the environment noise) is directed through a tunable loss module(with transmission \(\eta_e\), to simulate the additional loss of the probe channel, and including the detector efficiency as well) and detected by a second single photon detector. The total efficiency is given by \(\eta_p = \eta_e\eta_o\). The probe and idler photons detected on both detectors are time tagged for correlation analysis. In the ICC setup (shown in Fig.~\ref{ExpSetup}(b)), the correlated probe and reference photons (with same mean photon number per mode \(\mu\) ) are obtained through splitting the (locally) thermal state of the SPDC signal light on a balanced beamsplitter.\\

It must be noted that in our experimentally achievable implementation of the ICC protocol, the generated probe and reference photons are not optimal: the down-converted signal light are not in a single-mode thermal state, but rather a statistical mixture of thermal states of many orthogonal frequency modes. This spectral multimodeness could be characterized by the Schmidt number \(M\) of the down-converted states.  As shown in Eq. (4), the ICC coincidence counts is a function of the number of the Schmidt modes \(M\) of the down-converted photon pairs. The Schmidt number \(M\ge13\) for the photon pair source is obtained through the numerical Schmidt decomposition of the experimentally measured joint spectral intensity (Fig.~\ref{JSA}). By imposing \(M\) = 1, we obtain the best theoretical single-mode ICC protocol achievable with single-mode thermal state sources. This ideal case is used to calculate the quoted ratios  \(\epsilon_\text{ICQ}/\epsilon_\text{ICC}\) together with the experimental results from the ICQ protocol, to quantify the enhancement of the ICQ protocol.  \\

For each experimental data point, the contrast \(\epsilon_{ICQ}\) (\(\epsilon_{ICC}\)) of the ICQ (ICC) protocol are directly calculated according to the definition \eqref{SNR_DEF}, with experimentally measured photon counting statistics. To obtain a theoretical plot of \(\epsilon_\text{ICQ}\) and  \(\epsilon_\text{ICC}\) one can still apply this definition, but with photon counting statistics expressed as a function of the probe and reference efficiency, \(\eta_p\) and \(\eta_r\) respectively, and the pair generation rate of the source, \(\mu\) (full derivation of the photon counting statistics could be found in the appendix.)
\begin{align}
    \avg{N_p N_r}_\text{ICQ} &= \eta_p\eta_r \mu(\mu(1+\frac{1}{M})+1) + \eta_r\mu\avg{N_b}\label{NC} \\
\avg{N_p N_r}_\text{ICC} &= \eta_p\eta_r \mu^2(1+\frac{1}{M}) + \eta_r\mu\avg{N_b} \label{IC CIContrast} \\
    \avg{N_p} &= \eta_p \mu + \avg{N_b}\label{NP}\\
    \avg{N_r} &= \eta_r \mu\label{NR}\\
    \avg{\delta^2S} &= \avg{\delta^2(N_pN_r)} \simeq \avg{N_pN_r} -  \avg{N_pN_r}^2\label{temp}
\end{align}
where \(\avg{N_b}\) denotes the average noise photon number at the detector that overlap with each induividual probe pulse.
Note that the expressions \eqref{NC}-\eqref{temp} apply to both the target present (\(\eta_p=\eta_o\eta_e\)) and absent (\(\eta_p=0\)) case.
The values of \(\mu\), \(\eta_p\), and \(\eta_r\) used in the theoretical calculation are calculated from the averaged photon counting statistics from different experiments. 

\begin{figure}
\centering
(a)\\
\subfigure{\includegraphics[width=0.8\columnwidth]{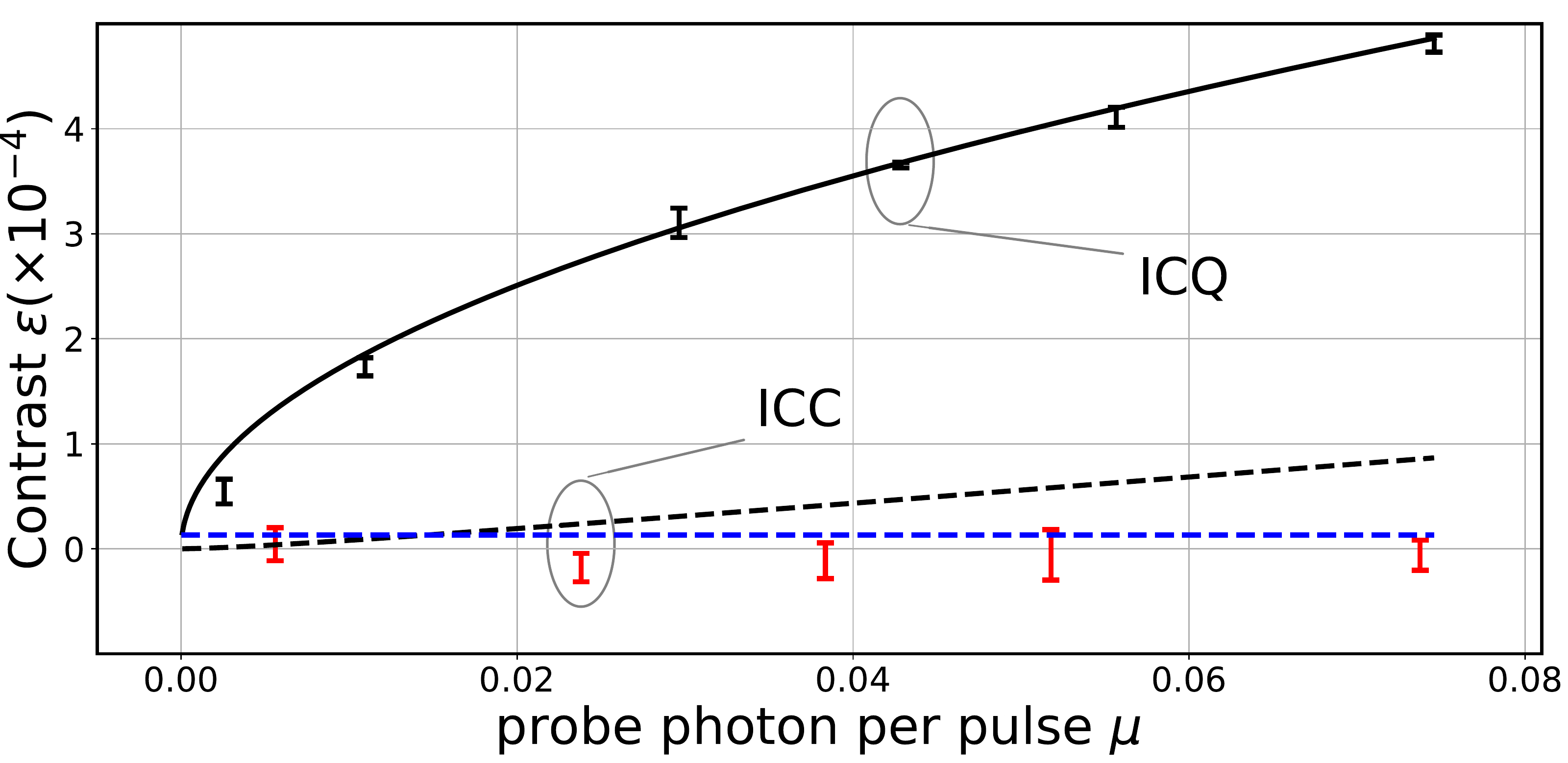}\label{PowerSNR}}\\
(b)\\
\subfigure{\includegraphics[width=0.8\columnwidth]{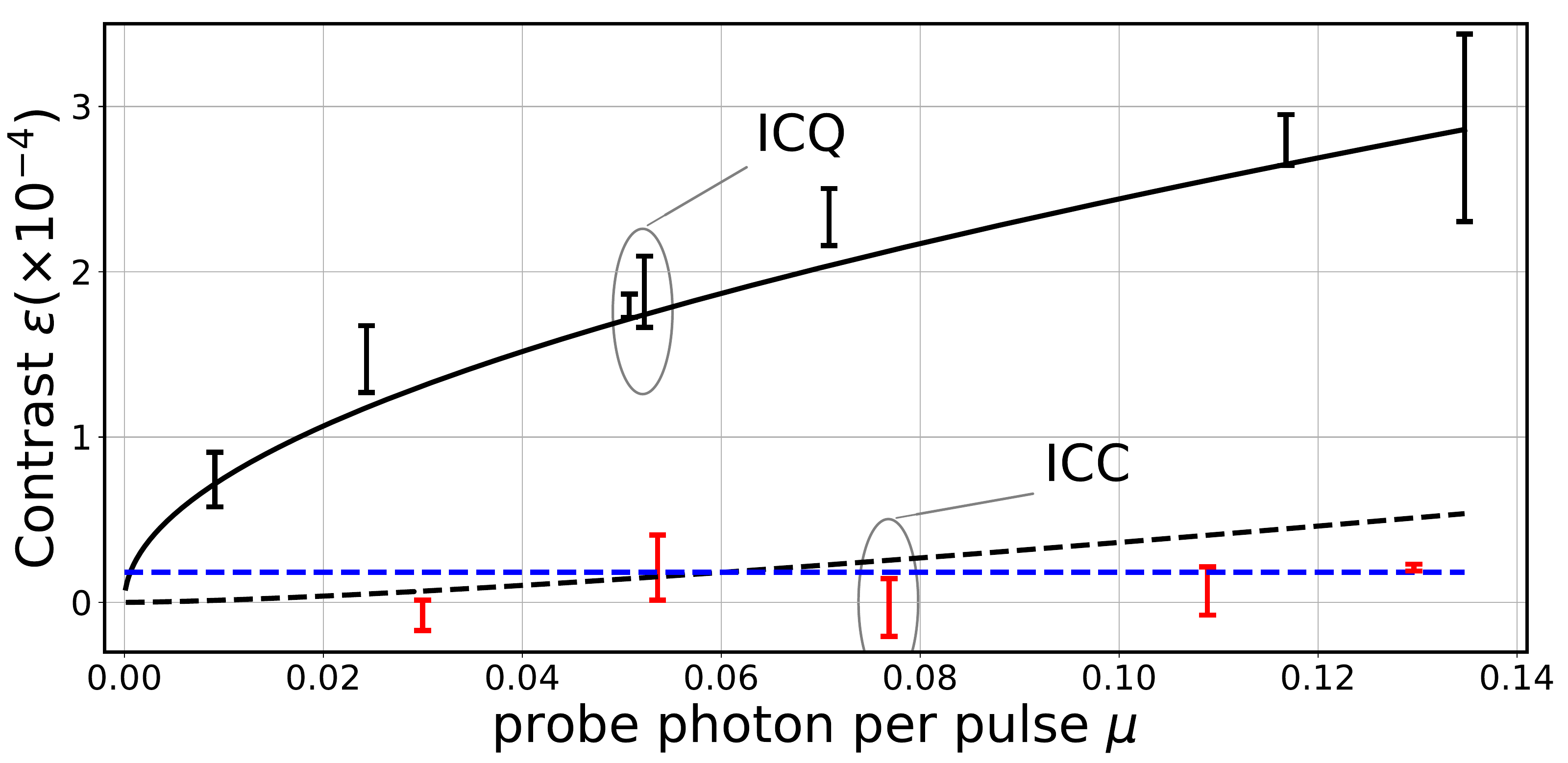} \label{PowerLossNoiseSNR}}\\
(c)\\
\subfigure{\includegraphics[width=0.85\columnwidth]{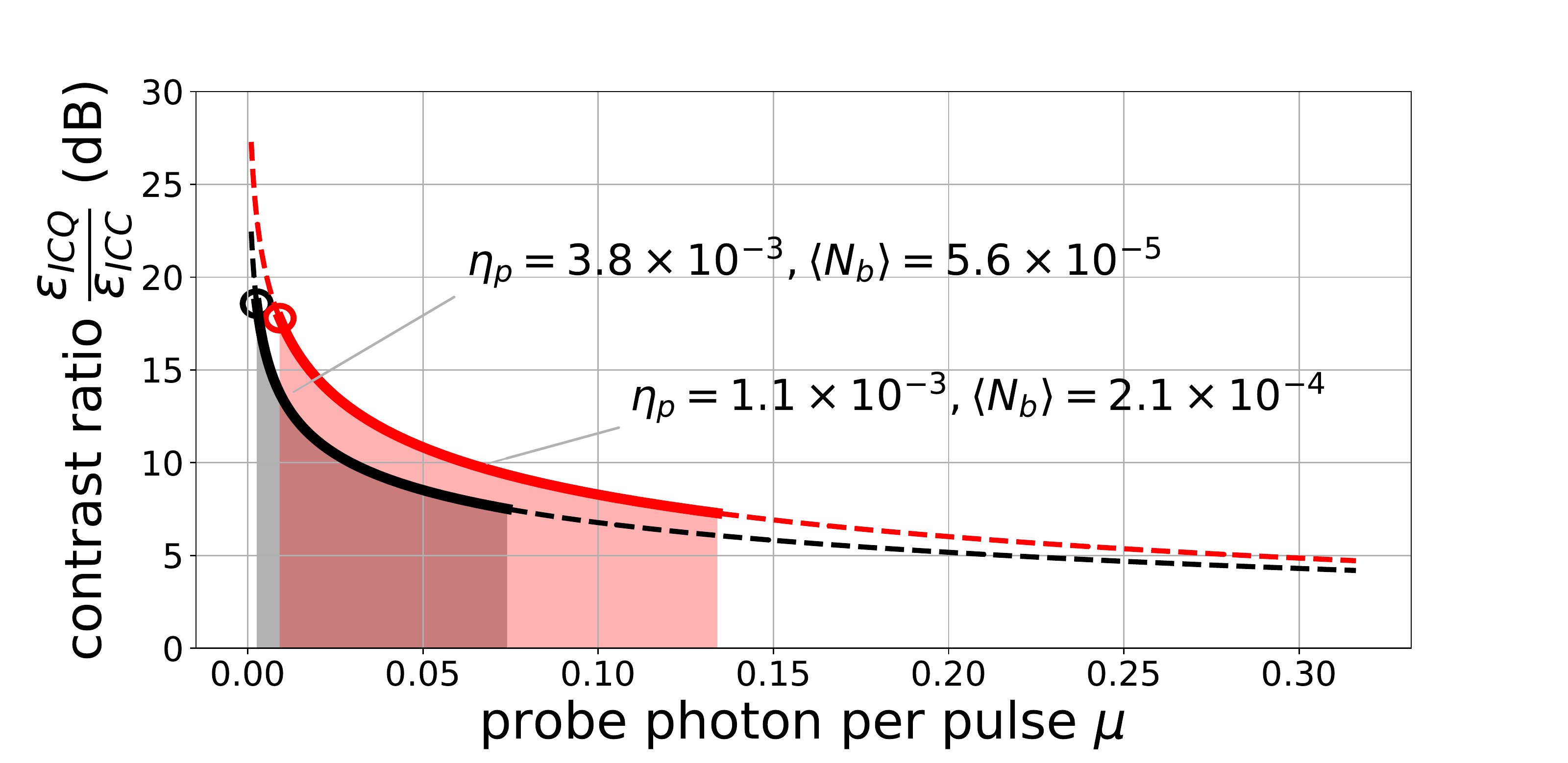} \label{PowerRatio}}
    \caption{\label{PowerPlots}\subref{PowerSNR} Normalized contrast \(\epsilon\) as a function of the average photon pair number generated per pulse, \(\mu\), for the ICQ (solid black) protocol, with no additional loss and noise(\(\eta_r = 8.6\times10^{-4},\eta_p=3.8\times10^{-3},\avg{N_b} = 5.6\times 10^{-5}\)). Black error bar: the experimentally measured contrast for the ICQ protocol. Dashed black curve: the bound of the maximum achievable contrast \(\epsilon\) for the single-mode ICC implementation (\(M=1\), with same channel transmission and noise power). Dashed blue line: the noise floor for the contrast of the ICC experiment. Error bars are given by the standard deviation of three measurements. \subref{PowerLossNoiseSNR} Normalized contrast \(\epsilon\) as a function of the average photon pair number generated per pulse, \(\mu\), with additional loss and noise injection(\(\eta_r = 7.3\times10^{-4},\eta_p=1.1\times10^{-3},\avg{N_b} = 2.1\times 10^{-4}\)). \subref{PowerRatio} Contrast ratio \(\epsilon_\text{ICQ}/\epsilon_\text{ICC}\) as a function of the average probe photon number generated per pulse \(\mu\) for plots \subref{PowerSNR} (black line) and \subref{PowerLossNoiseSNR} (red line). The value of \(\epsilon_\text{ICQ}\) and \(\epsilon_\text{ICC}\) are the theoretical value of contrast for the best ICC protocol and the ICQ protocol (the dashed and solid curve in (a) and (b)). The experimentally probed parameter region is marked by the shaded area for the two experiments. The circles on the left of the solid curve correspond to the maximal values of the \(\epsilon_\text{ICQ}/\epsilon_\text{ICC}\) ratio, being 18.57dB (without additional loss and noise) and 17.89db(with additional loss and noise) respectively.}
\end{figure}

\begin{figure}
\centering
(a)\\
\subfigure{\includegraphics[width=0.8\columnwidth]{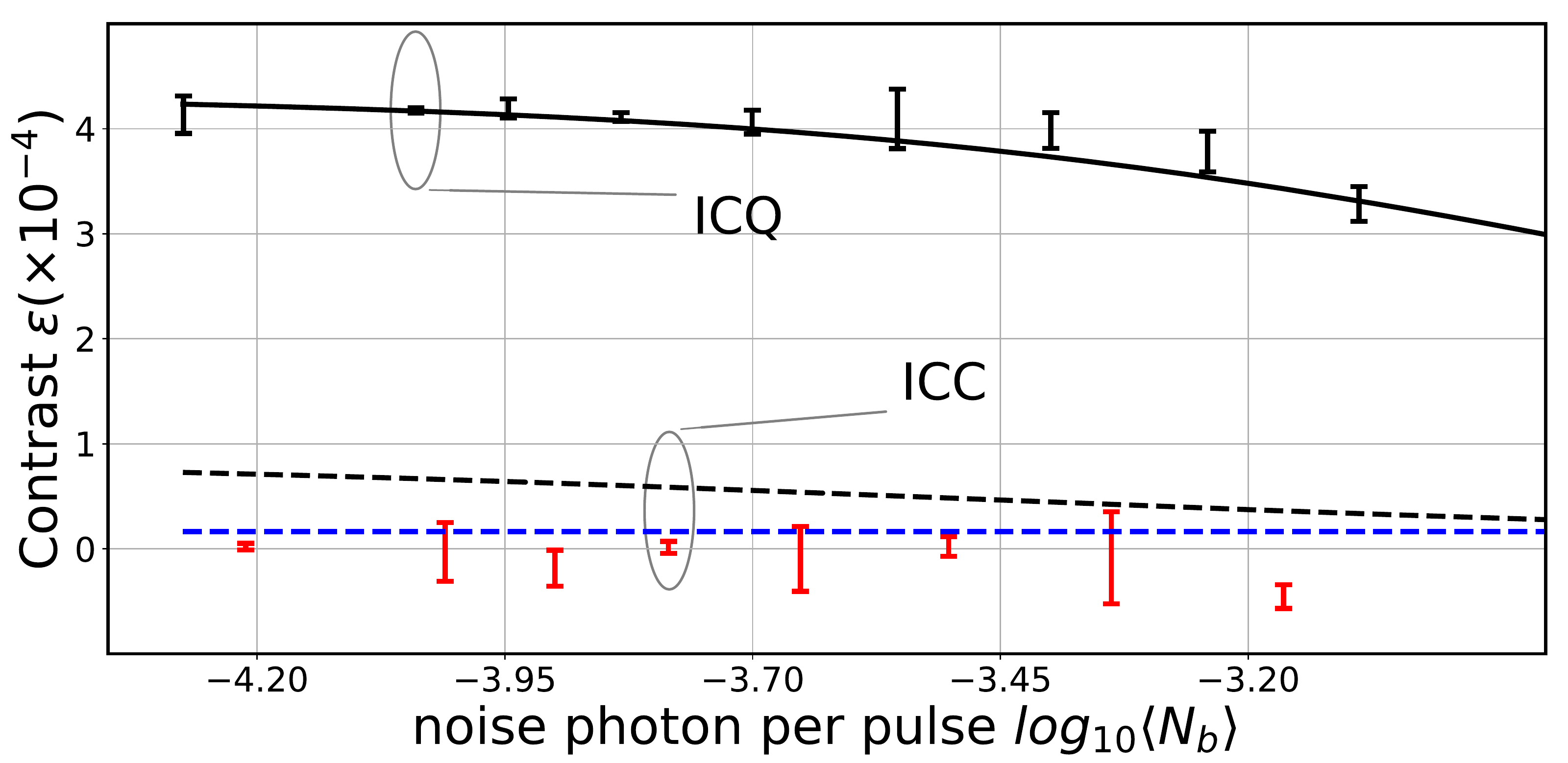} \label{NoiseSNR}}\\
(b)\\
\subfigure{\includegraphics[width=0.8\columnwidth]{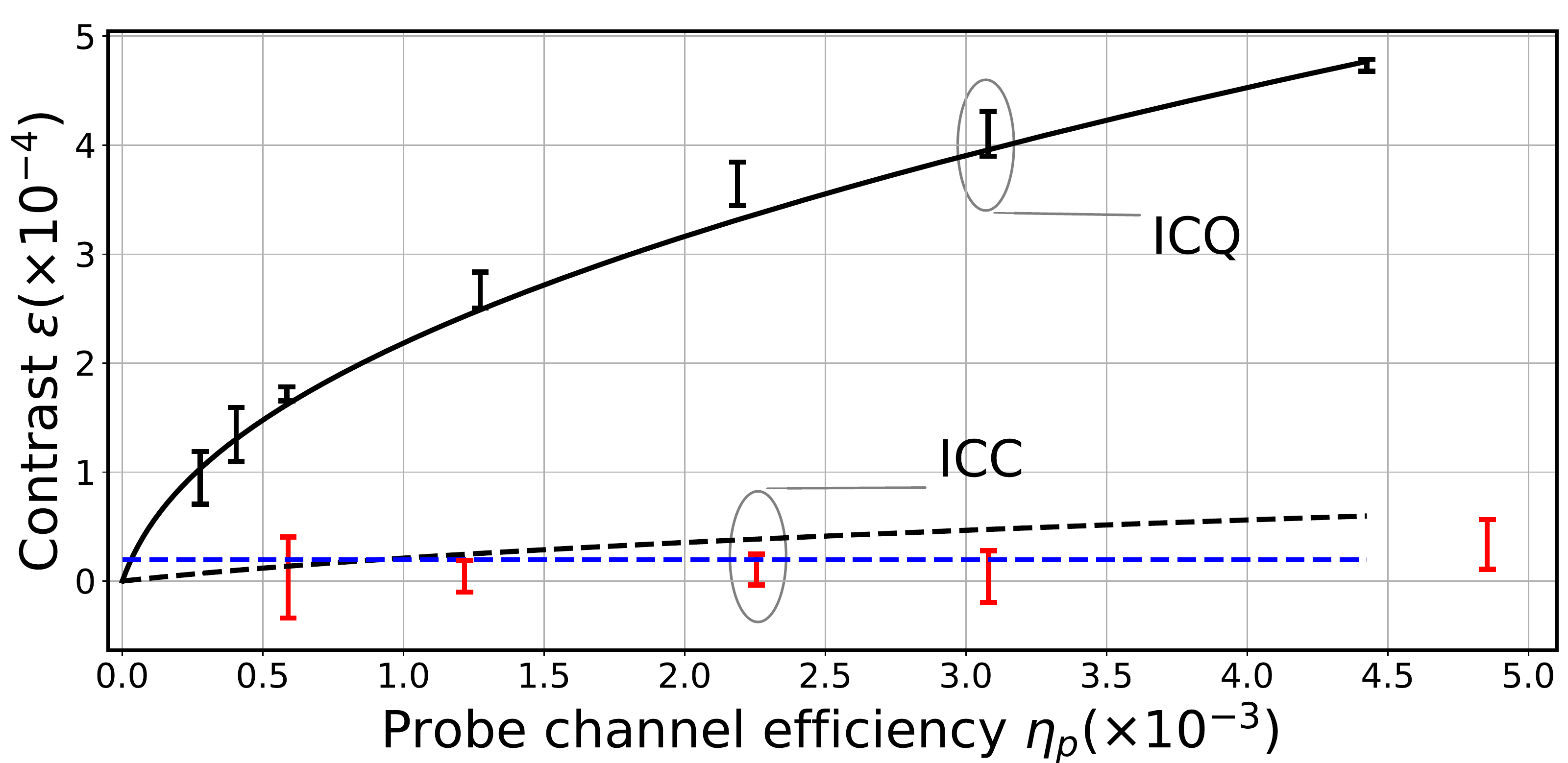} \label{LossSNR}}\\
(c)\\
\subfigure{\includegraphics[width=0.8\columnwidth]{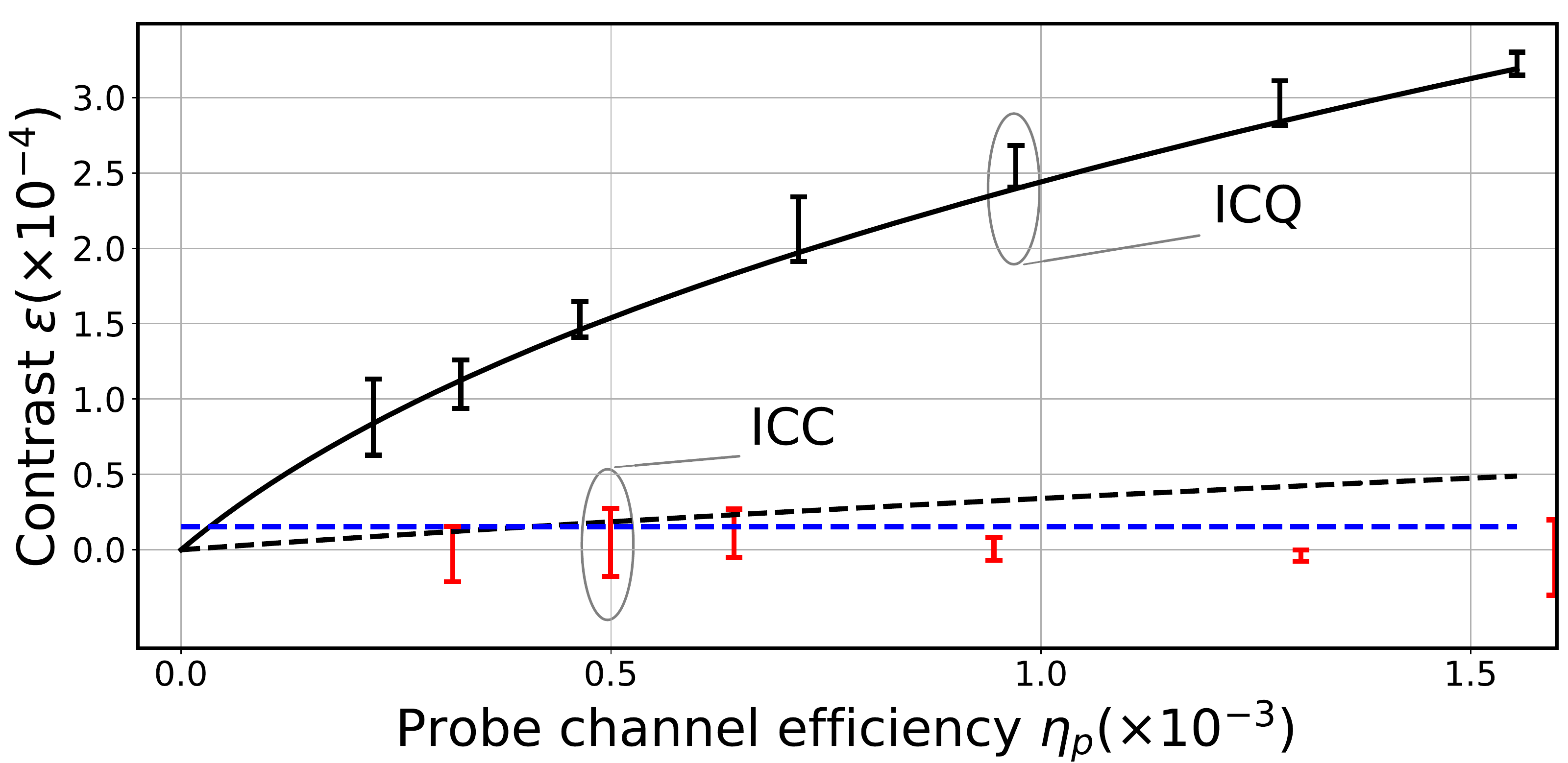} \label{LossNoiseSNR}}
    \caption{\label{NoiseLossPlots}\subref{NoiseSNR} Normalized contrast \(\epsilon\) as a function of the average noise photon number per pulse, \(\avg{N_b}\)( \(\eta_r=1.2\times10^{-3},\eta_p=2\times10^{-3}, \mu=0.077\) ). Black errorbar: the experimentally measured contrast for the ICQ protocol. Dashed black curve: the bound of the maximum achievable contrast \(\epsilon\) for the single-mode ICC implementation ( \(M=1\), with the same channel transmission and source pair rate). Dashed blue line: the noise floor for the contrast of the ICC experiment. The maximum \(\epsilon_\text{ICQ}/\epsilon_\text{ICC}\) achieved is \(\SI{9.95}{dB}\). \subref{LossSNR} Normalized contrast \(\epsilon\) as a function of the probe channel transmissivity with no additional noise (\(\eta_r=1.3\times 10^{-3},\mu=0.04,\avg{N_b} = 5.5\times10^{-5}\)). The maximum \(\epsilon_\text{ICQ}/\epsilon_\text{ICC}\) achieved is \(\SI{11.68}{dB}\). \subref{LossNoiseSNR} Normalized contrast \(\epsilon\) as a function of the probe channel transmission with additional noise,\(\eta_p\)(\(\eta_r=8.5\times10^{-4},\mu=0.093,\avg{N_b} = 2.05\times10^{-4}\)). The maximum \(\epsilon_\text{ICQ}/\epsilon_\text{ICC}\) achieved is \(\SI{9.82}{dB}\).}
\end{figure}

\section{Experimental Results}
Fig.~\ref{PowerPlots} shows a comparison between the ICQ and ICC protocol for different brightness levels, \(\mu\). In our experiment, the number of realizations is equal to the number of recorded pump pulses. The duration of each measurement is \(\SI{40}{s}\), with a number of valid realizations of around \(\num{2e9}\). The average probe photon number generated per pump pulse, \(\mu\), is calculated from the photon counting statistics(\eqref{NC},\eqref{NP},\eqref{NR}).
The ICQ protocol performance is compared to only the best theoretical (single-mode, \(M=1\)) implementation of ICC (``ICC bound'' in the plots). See the appendix for details of the experimental implementation and the effect of the Schmidt number of the photon pair source on the ICC contrast. \\

Under \(\SI{0}{dB}\) additional loss (introduced by the tunable loss module and the 50:50 beamsplitter) and \(\SI{0}{dB}\) additional noise (injected by the EDFA), the ICQ protocol shows an improvement of up to \(\SI{18.57}{dB}\) over the ICC contrast in low-brightness conditions for the region explored experimentally. This improvement is with respect to the theoretical ICC bound; the contrast ratio between the ICQ and ICC protocol becomes larger at smaller photon flux (lower \(\mu\)), highlighting the better performance for the ICQ protocol in low-brightness condition. When introducing additional loss (\SI{3}{dB} beamsplitter loss) and noise (\(\SI{13.40}{dB}\) compared to the detected probe signal), this quantum advantage is still found to be considerable, as shown in Figs.~\ref{PowerLossNoiseSNR} and \subref{PowerRatio}.
In both cases, the performance advantages of the ICQ protocol over the ICC protocol decrease as the source pair generation rate $\mu$ increases.

The performance of the ICQ and ICC protocol under different additional noise and loss levels are shown in Fig.~\ref{NoiseLossPlots}. The loss is varied by adjusting the coupling efficiency of the adjustable fiber optic attenuator. In both the ICQ and ICC experiment, the probe beam is mixed with thermal noise using a 50:50 fiber beamsplitter, introducing a \(\SI{3}{dB}\) loss into the probe channel. The ICQ and ICC contrast \(\epsilon\) as a function of the noise injected into the probe path, is shown in Fig.~\ref{NoiseSNR}. While the theoretical value of \(\epsilon\) for both ICQ and ICC decreases as more noise is injected into the probe path, ICQ is shown to be more resilient to noise compared to the ICC protocol. The ICQ advantage is also shown as a function of loss in the absence and presence of further noise, in Figs.~\ref{LossSNR} and \subref{LossNoiseSNR} respectively. This loss-tolerance property makes ICQ a suitable protocol to detect a low-reflection target in a high-loss environment.

\section{Discussion}
As evident from the results, a quantum enhancement has been demonstrated in the ICQ protocol using a monolithic, on-chip quantum light source. A contrast enhancement persists even in the presence of high levels of noise and additional loss in the channel. Our experimental protocol produces results comparable with previous experimental work in this area \cite{lopaeva2013experimental}.
 The ICQ protocol further shows its resilience to noise and loss, especially in the low-brightness regime. 
Since the detection of each probe and reference photon are time tagged, the photon counting statistics could also be used to calculate the traveled distance of the probe photons from the time of flight of the probe photon.  
In addition  to the performance enhancement, the compact semiconductor waveguide source of the ICQ protocol also enables large scale integration. \\

\begin{figure}[h]
\centering
\includegraphics[width=\columnwidth]{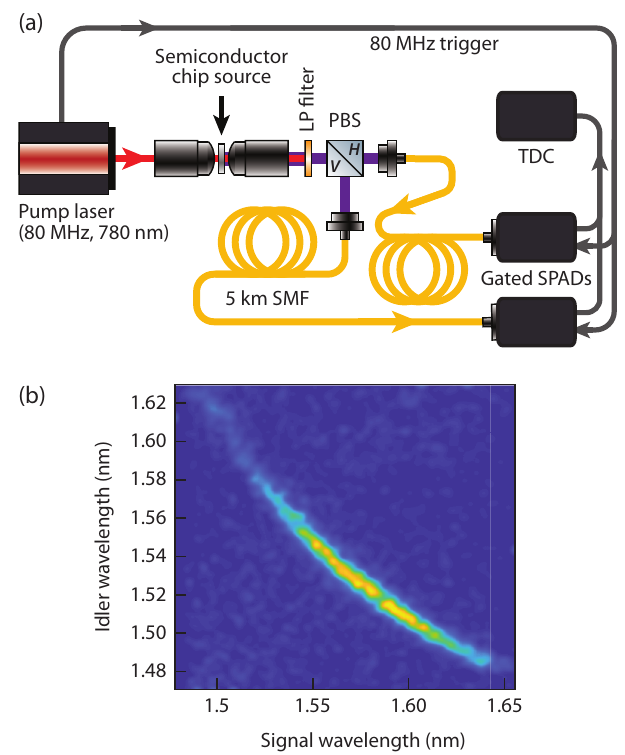}
\caption{(a) A schematic of the experimental setup for joint spectral intensity measurements using the frequency to time mapping technique. LP: long-pass filter (\(\ge1450\)nm); SMF: single-mode fiber; SPAD: single-photon avalanche diode; TDC: time-to-digital converted. (b) Joint spectral intensity result. The bandwidth of the SPDC photon is estimated to be around 100nm, which is limited by the long pass filter.}\label{JSA}
\end{figure}

A major limitation of the ICQ protocol is that the strength of intensity correlation between the probe and reference photon (hence the enhancement of the target detection performance) is limited by the average number of probe photon per pulse \(\mu\). When the mean photon number per pulse \(\mu\) decreases, the performance advantage of the ICQ protocol as compared to the ICC protocol increases, but at the price of sacrificing the absolute performance of the ICQ protocol, as could be seen in Fig. \ref{PowerSNR} and Fig. \ref{PowerRatio}. A possible approach to increasing the performance of the ICQ protocol without decreasing \(\mu\) is to utilize the frequency correlation that also exists within nonclassical photon pairs: while the frequency of the probe and reference photon are broadband individually, the sum of their frequencies could be within a narrow frequency range, which implies strong frequency correlations. To see how the frequency correlation could benefit target detection, it suffices to consider a photon pair state \(\ket{\psi}\) that is correlated in the frequency degree of freedom \(\ket{\psi}\):    
\begin{gather}
\ket{\psi} = \iint d\wp d\wr \psi^*(\wp,\wr) a_p\dg(\wp) a_r\dg(\wr)\ket{0}
\end{gather}
where \(a_p(\wp),a_r(\wr)\) are the annihilation operators of the probe and reference photon at frequency \(\wp\) and \(\wr\), respectively and the function \(\psi(\wr,\wr)\) is the joint spectral amplitude. It could be further shown that \(\ket{\psi}\) can be decomposed as a superposition of different photon pair states through the Schmidt decomposition:  
\begin{gather}
\ket{\psi} =\sum\limits_n \sqrt{\lambda_n}A_{p,n}\dg A_{r,n}\dg\ket{0}\label{SD}
\end{gather}
where \(\{A_{p,n}\},\{A_{r,n}\}\) are the different discrete frequency mode operators for the probe and reference light and \(\{\lambda_n\}\) are the corresponding Schmidt eigenvalues. The number of the mode pairs could be quantified as the Schmidt number \(M = 1/\sum\limits_n \lambda_n^2\). The expression \eqref{SD} suggests that, similar to the case of the ICQ protocol where probe and reference photons are always created in pairs in different pulse pairs, the frequency correlated probe and reference photons are also created in pairs in different frequency mode pairs \(\{(A_{p,n},A_{r,n})\}\). Therefore for each frequency mode pair \((A_{p,n},A_{r,n})\), the same intensity correlation analysis in the ICQ protocol applies. Since the number of different frequency mode pairs \(M\) could be very large,  each of the frequency mode pair \((A_{p,n},A_{r,n})\) can have mean photon number much less than one, which translates to high performance enhancement of the ICQ protocol. In general, the Schmidt number of the non-classical photon pair could be approximated by the ratio of the SPDC photon bandwidth and the SPDC pump bandwidth. For semiconductor waveguides with a specific structure design\cite{abolghasem2009bandwidth}, around 400nm of the SPDC photon bandwidth could be achieved. The detection of each frequency mode (\(A_{p,n}\) and \(A_{r,n}\)) could be done through frequency resolved photon counting or frequency to time mapping based on the fast fiber spectrogram technique\cite{avenhaus2009fiber}. Fig. \ref{JSA} shows the experimental setup and result of the measurement of the joint spectral intensity. Through numerical Schmidt decomposition of the joint spectral amplitude (which is assumed to be the square root of the joint spectral intensity), the Schmidt number of the pulsed SPDC photon pair source is estimated to be around \(M=13\). Larger Schmidt number \(M\) could be achieved with a narrowband pump.

\section{Conclusions}
We demonstrated an intensity correlation target detection protocol enhanced by non-classical light generated in a semiconductor chip source. This is the first instance where a quantum enhancement in a target detection protocol over a thermal background has been shown in an integrated platform. Our device can achieve up to \(\SI{18.57}{dB}\) experimentally verified contrast improvement over the classical intensity correlation target detection protocol, in the absence of additional loss and noise. A high quantum contrast has also been measured even under both \(\SI{29.69}{dB}\) loss, and noise \(\SI{13.40}{dB}\) stronger than the detected probe field. The ratio between the experimental value of \(\epsilon_\text{ICQ}\) and the best theoretical \(\epsilon_\text{ICC}\) in equivalent conditions is \(\SI{17.79}{dB}\). 
When separately analyzing the system performance in terms of noise and additional loss, we have experimentally demonstrated a contrast enhancement up to \(\SI{9.95}{dB}\) as a function of noise, and \(\SI{11.68}{dB}\) as a function of loss. We also proposed a method to further improve the performance of the ICQ protocol by utilizing the strong frequency correlations of SPDC photon pairs.


%

\appendices
\section{Experiment Setup Details}

We refer to the photons in different polarization generated in the semiconductor waveguide as signal and idler photons. For the ICQ experiment, the signal photon is used as a ``local'' reference and detected by an MPD InGaAs single-photon detection module, gated by the pulsed pump. The idler photon is used as the probe and mixed on a 50:50 fiber beamsplitter with a broadband amplified spontaneous emission (ASE) noise produced by an erbium-doped fiber amplifier (EDFA). The ASE noise level is adjusted through a variable attenuator. The mixed output is detected by an id210 single-photon detector. To simulate the absence of a target, a beam block is placed in the idler path. The id210 is externally triggered by the \(\SI{80}{MHz}\) pump, open for \(\SI{3}{ns}\), with \(\SI{20}{\micro\second}\) dead time, and quantum efficiency set to \(\SI{25}{\percent}\). The time-to-digital converter used to correlate the signal is an id800 model.

For the ICC  experiment, only the signal path is used, with the idler path blocked. The signal beam passes through a 50:50 fiber beamsplitter to produce correlated thermal light. One path is used as a reference path and detected by the MPD detector. The probe path is mixed with ASE noise by a 50:50 fiber beamsplitter and detected by the id220 detector. Tunable losses are introduced by adjusting the fiber-to-fiber coupling in the probe path. The no-target condition is simulated by disconnecting the probe path fiber from the fiber beamsplitter input.

\renewcommand{\v}[1]{\ensuremath{\mathbf{#1}}} 
\section{Derivation of photon counting statistics}
In the experiment, the light source used are Gaussian states and the channel loss and noise could be modeled as Gaussian operation upon Gaussian states\cite{olivares2012quantum}. As a result, the Gaussian state formalism could be used to compute the expectation value (3)-(6). The advantages of using Gaussian state formalism to compute (3)-(6) is two-fold:first it does not require complicated operator expansion (e.g. unitary transform between the input and output modes of the beam-splitter) so the computation is much more scalable as the complexity of the optical setup increase; secondly the evolution of the quantum states could be modeled as the transform of their covariance matrices and could be calculated symbolically with program. A Gaussian state is completely characterized by its covariant matrix and first-order moment:
\begin{gather}
    \sigma_{kl} = \frac{1}{2}\avg{\hat{R}_k\hat{R}_l+\hat{R}_k\hat{R}_l} - \avg{\hat{R}_k}\avg{\hat{R}_l}\\
    d_l = \avg{\hat{R}_l}
\end{gather}
Where \(\hat{R}_{2i-1},\hat{R}_{2i}\) is the \(x\) and \(p\) quadrature of the \(i\)th mode, respectively. The commutation relationship could be expressed:
\begin{gather}
    [\hat{R}_k , \hat{R}_l] = i\Omega_{kl}\\
    \Omega_{kl} = \otimes_{i=0}^n\omega\\
    \omega = 
    \begin{bmatrix}
    &0&1\\
    &-1&0\\
    \end{bmatrix}
\end{gather}
Throughout the experiment setup, all the quantum states have zero first order moment (no coherent state component). The covariance matrices of the down-converted photon pairs source and the correlated thermal light source is given by:
\begin{widetext}
\begin{gather}
    \sigma_{SPDC} = 
    \begin{bmatrix}
        &\mu+\frac{1}{2}&0&\sqrt{\mu(\mu+1)}&0\\
        &0&\mu+\frac{1}{2}&0&\sqrt{\mu(\mu+1)}\\
        &\sqrt{\mu(\mu+1)}&0&\mu+\frac{1}{2}&0\\
        &0&\sqrt{\mu(\mu+1)}&0&\mu+\frac{1}{2}\\
    \end{bmatrix}\\
    \sigma_{thermal} = 
    \begin{bmatrix}
        &\mu+\frac{1}{2}&0&\mu&0\\
        &0&\mu+\frac{1}{2}&0&\mu\\
        &\mu&0&\mu+\frac{1}{2}&0\\
        &0&\mu&0&\mu+\frac{1}{2}\\
    \end{bmatrix}
\end{gather}
\end{widetext}
To model the loss on each channel, one could first append (direct sum) a vacuum covariance matrix:
\begin{gather}
  \sigma_{vac} = \frac{I_{2\times2}}{2}   
\end{gather}
to the covariance matrix of the source and then apply a symplectic transform that corresponds to beam-splitting(with power transmission \(\eta\)) on the joint covariance matrix :
\begin{gather}
    S_{bs} = 
    \begin{bmatrix}
        &\sqrt{\eta}&0&\sqrt{1-\eta}&0\\
        &0&\sqrt{\eta}&0&\sqrt{1-\eta}\\
             &-\sqrt{1-\eta}&0&\sqrt{\eta}&0\\
                  &0&-\sqrt{1-\eta}&0&\sqrt{\eta}\\
    \end{bmatrix}
\end{gather}
(note that only the relevant dimensions are shown in the symplectic transform. The other dimensions of the symplectic transform corresponding to non-interacting modes are identities and omitted)Then the covariant matrix after the loss is given by:
\begin{gather}
    \sigma_{loss} = S_{bs}\sigma_{source}S_{bs}^T
\end{gather}
where \(\sigma_{source}\) could be either \(\sigma_{SPDC}\) for ICQ or \(\sigma_{thermal}\) for ICC.  Mixing with thermal noise could be treated similarly, except that a thermal noise state with covariance matrix:
\begin{gather}
    \sigma_{noise} = I_{2\times2}\avg{N_b}+\sigma_{vac}
\end{gather}
 is appended(direct sum) to the source covariant matrix instead and the beam-splitting is 50-50(\(\eta = 50\%\)). After going through the loss and mixing with noise and tracing out the unused ouput mode of the beam-splitter, the final covariance matrices \(\sigma_{SPDC,fin}\) \(\sigma_{thermal,fin}\) is given by: 
\begin{widetext}
\begin{gather}
    \sigma_{SPDC,fin} = 
    \begin{bmatrix}
        &\avg{N_b}+\eta_p\mu+\frac{1}{2}&0&\sqrt{\eta_p\eta_r\mu(\mu+1)}&0\\
        &0&\avg{N_b}+\eta_p\mu+\frac{1}{2}&0&\sqrt{\eta_p\eta_r\mu(\mu+1)}\\
        &\sqrt{\eta_p\eta_r\mu(\mu+1)}&0&\eta_r\mu+\frac{1}{2}&0\\
        &0&\sqrt{\eta_p\eta_r\mu(\mu+1)}&0&\eta_r\mu+\frac{1}{2}\\
    \end{bmatrix}\\
    \sigma_{thermal,fin} = 
    \begin{bmatrix}
        &\avg{N_b}+\eta_p\mu+\frac{1}{2}&0&\sqrt{\eta_p\eta_r\mu^2}&0\\
        &0&\avg{N_b}+\eta_p\mu+\frac{1}{2}&0&\sqrt{\eta_p\eta_r\mu^2}\\
        &\sqrt{\eta_p\eta_r\mu^2}&0&\eta_r\mu+\frac{1}{2}&0\\
        &0&\sqrt{\eta_p\eta_r\mu^2}&0&\eta_r\mu+\frac{1}{2}\\
    \end{bmatrix}
\end{gather}
\end{widetext}
The final state is completely specified by the covariance matrices and the expectation value of the left-hand side of equation (3)(4)(5)(6) could be calculated according to \cite{cahill1969ordered} and shown to be:
\begin{align}
    \avg{N_p N_r}_\text{ICQ} &= \eta_p\eta_r \mu(2\mu+1) + \eta_r\mu\avg{N_b} \label{NC_old} \\
            \avg{N_p N_r}_\text{ICC} &= 2\eta_p\eta_r \mu^2 + \eta_r\mu\avg{N_b} \label{NC_ci_old} \\
        \avg{N_p} &= \eta_p \mu + \avg{N_b}\label{NP_old}\\
        \avg{N_r} &= \eta_r \mu\label{NR_old},
\end{align}
To calculate the higher order moment \(\avg{\delta^2S}\):
\begin{gather}
            \avg{\delta^2S}\\
	=\avg{ \delta^2(N_rN_p-\avg{N_r}\avg{N_p})}\\
	=\avg{ \delta^2(N_rN_p)}\\
	=\avg{ N_rN_pN_rN_p - \avg{N_rN_p}^2}\\
	\simeq\avg{ N_rN_p} - \avg{N_rN_p}^2
\end{gather}
The approximation in the last equation is valid becasue in the intensity regime we are interested in, operator \(N_rN_p\) only have eigenvalue 0 and 1. Thus \(N_rN_pN_rN_p = N_rN_p\).\\

In the analysis above single-mode SPDC/thermal source is assumed. However, in the actual experiment, the SPDC state generated by a single pump pulse is a tensor product of many simultaneously squeezed states on different probe/reference mode pairs, and the state obtained by splitting the H polarized SPDC photons(which are generated within a single pump pulse) is a tensor product of many correlated thermal state pairs:
\begin{gather}
    \rho_{\text{multi\_SPDC}}   = \otimes_{n=1}^M \rho_\text{SQZ}\\
    \rho_{\text{multi\_thermal}} = \otimes_{n=1}^M \rho_\text{thermal\_pair}
\end{gather}
Experimentally, the singles counting on each channel is to count the total number of probe/reference photons in all the M probe/reference modes and could be model mathematically as:
\begin{gather}
    \avg{N_p} = \sum\limits_{n=1}^M \avg{N_{p,n}}\\
    \avg{N_r} = \sum\limits_{n=1}^M \avg{N_{r,n}}
\end{gather}
Where \(N_{p,n}\) and \(N_{r,n}\) are the photon number operator of the probe/reference mode  of the \(n\)th probe-reference mode pair.Similarly, the coincidence counting is modeled by the total number of probe photons in all the \(M\) probe modes times the total number of the reference photon in all the \(M\) reference modes.
\begin{gather}
    \avg{N_pN_r} = \avg{\sum\limits_{h=1}^MN_{p,h}\sum\limits_{k=1}^MN_{r,k}}\\
    =\sum\limits_{h\ne k}^M\avg{N_{p,h}}\avg{N_{r,k}}\\
    +\sum\limits_{n=1}^M\avg{N_{p,n}N_{r,n}}
\end{gather}
For simplicity, we assume the total \(\mu\) probe/reference photons and is evenly distributed among all \(M\) probe/reference modes and the total \(\avg{N_b}\) noise photon are evenly distributed among the \(M\) probe modes. Then for each of the \(M\) probe/reference pair, the signal mode result \eqref{NC}-\eqref{NR}  applies  with \(\mu\rightarrow\frac{\mu}{M}\) and \(\avg{N_b}\rightarrow\frac{\avg{N_b}}{M}\):
\begin{align}
    \avg{N_{p,n} N_{r,n}}_\text{ICQ} &= \eta_p\eta_r \frac{\mu}{M}(2\frac{\mu}{M}+1) + \eta_r\frac{\mu}{M}\frac{\avg{N_b}}{M}  \\
        \avg{N_{p,n} N_{r,n}}_\text{ICC} &= 2\eta_p\eta_r (\frac{\mu}{M})^2 + \eta_r\frac{\mu}{M}\frac{\avg{N_b}}{M}  \label{ICCContrast} \\
        \avg{N_{p,n}} &= \eta_p \frac{\mu}{M} + \frac{\avg{N_b}}{M}\\
        \avg{N_{r,n}} &= \eta_r \frac{\mu}{M}
\end{align}
Then the experimentally measured photon counting statistics becomes:
\begin{gather}
    \avg{N_pN_r}_\text{ICQ} = \eta_p\eta_r(\mu+\mu^2+\frac{\mu^2}{M})+\eta_r\avg{N_b}\mu\label{exact}\\
    \avg{N_pN_r}_\text{ICC} = \eta_p\eta_r\mu^2(1+\frac{1}{M})+\eta_r\avg{N_b}\mu\\
    \avg{N_p} = \eta_p\mu+\avg{N_b}\\
    \avg{N_r} = \eta_r\mu
\end{gather}
It could be seen that the coincidence counting in ICQ case is only slightly affected by the multimodal nature of the source because in typical SPDC regime \(\mu\gg\frac{\mu^2}{M}\). Thus for simplicity,\eqref{NC_old}  could still be used instead of \eqref{exact}   as a good approximation.
\section{comparison with intensity detection protocol}
To give a more complete assessment of the ICQ protocol, we also compare its performance to classical target detection protocols with coherent probe light and intensity detection. To be specific, we will consider the following two types of classical intensity detection protocol for comparison:
\begin{itemize}
\item (a) the classical coherent state light source transmits a single probe light pulse that contains a large number of photons. 
\item (b) the classical coherent state light source transmits a train of probe light pulses, each contains the same average number of photons as the ICQ protocol.
\end{itemize}

To quantify the performance of the protocol (a), it suffices to calculate its error probability \(P_{e,COH}\):
\begin{gather}
P_{e,COH} \simeq  \exp(- \mu_{coh}\eta_p )/2
\end{gather}
where \(\mu_{coh}\) is the total number of photons in the probe pulse. To reach the error probability level of \(10^{-4}\), the number of detected photons \(\eta_p\mu_{coh}\) needs to reach 8.5. Note that the effect of background noise is negligible (\(\avg{N_b}=2.05\times10^{-4}\)), because of the small overlap of the single probe pulse and the environmental background noise. On the other hand, the error probability of the ICQ protocol is approximately given by:  
\begin{gather}
P_{e,ICQ} = \frac{1}{\sqrt{2\pi}}\int_{\sqrt{K}\epsilon'_{ICQ}}^\infty\exp(-t^2/2)dt
\end{gather}
where \(K\) is the number of transmitted probe pulses and \(\epsilon'_{ICQ}\) is the modified contrast defined as: 
\begin{gather}
\epsilon'_{ICQ} = \frac{\avg{S_\text{in}}-\avg{S_\text{out}}}{\sqrt{\avg{\delta^2 S_\text{in}}}+\sqrt{\avg{\delta^2 S_\text{out}}}}
\end{gather}
Assuming \(\mu=0.04, \eta_p = 3.8\times10^{-3}, \avg{N_b}=2.05\times{10^{-4}},\eta_r = 100\%\), then the number of probe pulses \(K\) that is required to achieve \(P_{e,ICQ}=10^{-4}\) is \(1.31\times10^5\), which correspond to detection of \(\eta_p\mu K = 20\) probe photons. The reason for the inferior performance of the ICQ protocol compared to the protocol (a) is that the concentration of the probe light in one single pulse can effectively reduce the overlap between the probe light with the continuous wave background noise. Therefore in the protocol (a) the effect of the noise background on the target detection performance is minimized.\\
Although sending a bright single probe pulse can reduce the effect of environmental noise on the target detection performance, it is still desirable to spread the probe light into multiple pulses in some target detection scenario, such as stealth operation, where the distinguishability between the probe photons and the CW background noise is to be minimized. This corresponds to the protocol (b), whose performance can also be quantified in the form of contrast similar to \eqref{SNR_DEF}:
\begin{gather}
\epsilon_{INT} = \frac{\avg{N_p}_{in}-\avg{N_p}_{out}}{\sqrt{\avg{N_p}_{in}+\avg{N_p}_{out}}}
\end{gather}
Note that for coherent state probe light the variance of the photon number equals the mean value of the photon number. The figure below compares \(\epsilon_{ICQ}\), \(\epsilon_{ICC}\) and \(\epsilon_{INT}\) as functions of the environmental noise power \(\avg{N_b}\) as well as the transmission of the reference photons \(\eta_r\). It could be seen that (1) in the limit of perfect reference photon transmission \(\eta_r=1\), the ICQ protocol has non-trivial performance advantage compared to protocol (b) and (2) the performance of the ICQ protocol is closely related to the transmission of the reference photon.
\begin{figure}
\centering
\includegraphics[width=0.8\columnwidth]{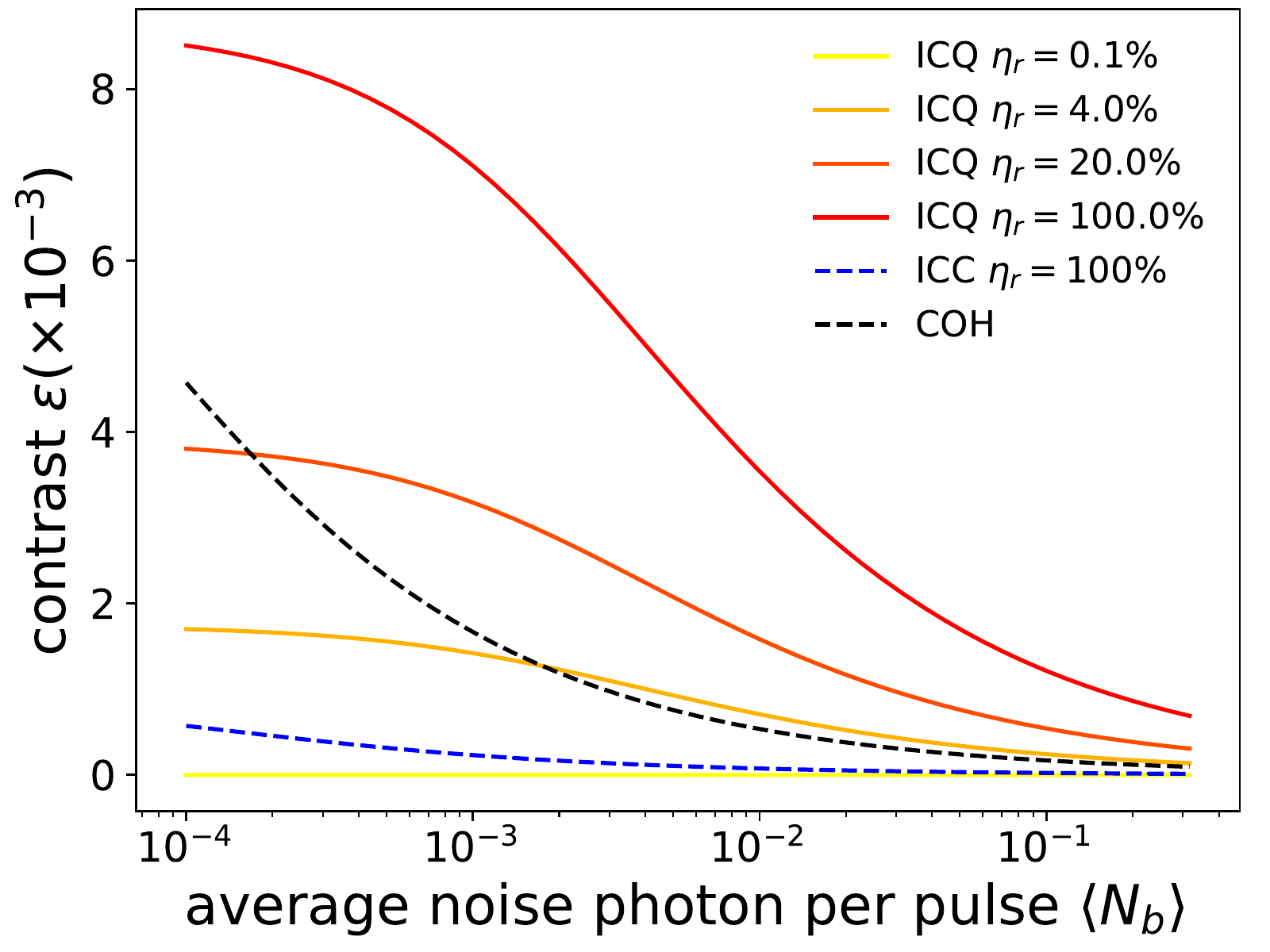}
\caption{The contrast of the ICQ, ICC and intensity detection protocol as a function of the environmental noise power \(\avg{N_b}\). The transmission of the probe and the probe power is given by \(\mu=3.8\times10^{-2}\) and \(\eta_p = 3.8\times10^{-3}\). }
\end{figure}

The authors thank OptoElectronic Components for the loan of an MPD InGaAs single-photon detection module.

\ifCLASSOPTIONcaptionsoff
  \newpage
\fi



\bibliographystyle{IEEEtran}
\bibliography{References}
%

%




\end{document}